\documentclass[sigconf]{acmart}

\usepackage{amsmath}               
  {
      \theoremstyle{plain}
      \newtheorem{assumption}{Assumption}
  }

\begin{document}

\copyrightyear{2018} 
\acmYear{2018} 
\setcopyright{acmcopyright}
\acmConference[KDD '18]{The 24th ACM SIGKDD International Conference on Knowledge Discovery \& Data Mining}{August 19--23, 2018}{London, United Kingdom}
\acmBooktitle{KDD '18: The 24th ACM SIGKDD International Conference on Knowledge Discovery \& Data Mining, August 19--23, 2018, London, United Kingdom}
\acmPrice{15.00}
\acmDOI{10.1145/3219819.3219860}
\acmISBN{978-1-4503-5552-0/18/08}

\title{False Discovery Rate Controlled Heterogeneous Treatment Effect Detection for Online Controlled Experiments}

\author{Yuxiang Xie}
\authornote{Work mostly done when the first author was interning at Snap Inc.}
\affiliation{%
  \institution{Department of Biostatistics}
  \institution{University of Washington}
  \city{Seattle, WA 98195}}
\email{yxxie@uw.edu}

\author{Nanyu Chen}
\authornote{Work done when this author was at Snap Inc.}
\affiliation{%
  \institution{LinkedIn}
  \institution{1000 W Maude Ave}
  \city{Sunnyvale, CA 94085}}
\email{nchen@linkedin.com}

\author{Xiaolin Shi}
\affiliation{%
  \institution{Snap Inc.} 
  \institution{63 Market St}
  \city{Venice, CA 90291}}
\email{xiaolin@snap.com}

\begin{abstract}
    Online controlled experiments (a.k.a. A/B testing) have been used as the mantra for data-driven decision making on feature changing and product shipping in many Internet companies. However, it is still a great challenge to systematically measure how every code or feature change impacts millions of users with great heterogeneity (e.g. countries, ages, devices). The most commonly used A/B testing framework in many companies is based on Average Treatment Effect (ATE), which cannot detect the heterogeneity of treatment effect on users with different characteristics. In this paper, we propose statistical methods that can systematically and accurately identify Heterogeneous Treatment Effect (HTE) of any user cohort of interest (e.g. mobile device type, country), and determine which factors (e.g. age, gender) of users contribute to the heterogeneity of the treatment effect in an A/B test. By applying these methods on both simulation data and real-world experimentation data, we show how they work robustly with controlled low False Discover Rate (FDR), and at the same time, provides us with useful insights about the heterogeneity of identified user groups. We have deployed a toolkit based on these methods, and have used it to measure the Heterogeneous Treatment Effect of many A/B tests at Snap.

\end{abstract}


\maketitle

\section{Introduction}

%
%
%
%

A/B testing, also known as controlled experiment \citep{Kohavi2016}, has become a common practice for evaluating and improving new product ideas across internet companies. Many IT companies with rich and large-scale data have built in-house A/B
testing platforms to meet their complex experimentation needs \citep{kohavi13, xu15}.
Some have been discussed at length in past technical papers such as \cite{kohavi13, xu15, xu16}, 
including the best practices and pitfalls \citep{deng17, Kohavi2012}.

At Snap, we have seen significant growth in A/B
test usage over the past two years. The in-house platform now hosts hundreds of concurrent experiments at any given time. Each experiment provides automated results for hundreds to thousands of diverse online metrics, ranging from engagement metrics, app performance metrics, to metrics used to understand data quality problems such as missing or duplicated event logging. 

As grown experimentation popularity comes with
greater needs, experimenters are no longer satisfied with knowing which metrics are impacted overall in an A/B test experiment, but have also become interested in knowing "why" metrics move or "who" drives the changes. Such insights about user heterogeneity can always help experimenters come up with strategies to improve the product. As an example, in a recent experiment we ran, we found that a metric decline was driven by users with the most snap views. With that observation we focused on understanding engineering and design behaviors when a user has large snap stacks to load, and were able to identify a critical performance issue which led to the metric decline. In fact, we have seen many cases where our users react differently to the same experiment treatment. For example, an experiment involving a minor feature adjustment received very
different feedback from users in different countries, with the same engagement metric increasing in some and decreasing in others. Another example was that code changes to improve app performance resulted in mixed outcomes across different mobile devices. Thus, detecting such heterogeneity of treatment effects in online controlled experiments has received great needs in our daily practice of using A/B testing to make decisions on product changes.

On the other hand, with the overly affluent amount of data, there is a strong threat from false discoveries, largely due to a statistical artifact known as "multiple testing"\footnote{https://en.wikipedia.org/wiki/Multiple\_comparisons\_problem}. With the hundreds of thousands of user characteristics available to internet companies, one can construct user groups in millions of ways. If we take a "naive" approach by simply computing and comparing the estimated effect based on users within groups, we can always easily find groups with treatment effects that differ widely from the average population, regardless of whether there is a real heterogeneity or not.

The goal of our work is to fill such gap by providing rigorous statistical methods and a toolkit that can detect Heterogeneous Treatment Effect (HTE) while dealing with the potential multiple testing problem by controlling the false positive rate (FDR) \footnote{https://en.wikipedia.org/wiki/False\_discovery\_rate}. This toolkit has been deployed and used at Snap. In this paper, we discuss the rationale behind using FDR and compare two statistical methods that control FDR using simulated results as well as real experimentation data. Depending on the methods we choose, we will discuss solutions to two questions experimenters and practitioners are interested in regarding HTE: 
\begin{itemize}
    \item Systematically find out which subgroups (e.g. countries) of users have treatment effect significantly different from the Average Treatment Effect of an A/B test.
    
    \item Rigorously figure out which factors (e.g. age, gender, etc.) contribute to the heterogeneity of the treatment effect in an A/B test.
\end{itemize}

Here is a summary of our contributions in this paper: 

\begin{itemize}
    \item We establish the HTE detection problem as an FDR control problem, and discuss in details why FDR control is important in large-scale HTE detection problem in practice.
    \item We apply two methods that can control FDR to our HTE detection process, and share insightful comparisons of the two methods based on simulation and real-world empirical data. 
    \item We share discussions on two critical lessons we learned, regarding (1) difference of heterogeneity in population v.s. heterogeneity in treatment effects; and (2) scalability of the algorithms. These insights should be able to help practitioners avoid pitfalls alike.
\end{itemize}

\section{Preliminaries}
\subsection{Average Treatment Effect vs. Heterogeneous Treatment Effect}

In an A/B test, we randomly split users into a treatment group and a control group, and observe the metrics of interest for all the users. The Rubin Causal Model \cite{holland86} is commonly used in A/B testing as a statistical framework for causal inference. Define $Y_i(T_i)$ to be the potential outcome for $i$-th user, where $T_i=1$ if the $i$-th user is in the treatment group and $T_i=0$ if the $i$-th user is in the control group. Therefore, $\tau_i = Y_i(1)-Y_i(0)$ is the causal effect of taking the treatment for $i$-th unit, and the average causal effect of all users $\Bar{\tau}$ is defined as the Average Treatment Effect (ATE). Note that ATE is not observable since we do not know $Y_i(0)$ and $Y_i(1)$ at the same time. This is known as the "fundamental problem of causal inference" \citep{holland86}. However, The estimator 
\begin{equation}
\overline{Y_{i|T_i = 1}} - \overline{Y_{i|T_i = 0}}
\end{equation}
is unbiased for ATE when the following two assumptions hold and is usually used for estimating ATE in an A/B test.

\begin{assumption}{Stable unit treatment value assumption (SUTVA):} \label{assump1}
\begin{itemize}
    \item Only one version of treatment and control, i.e. only one version of  $T=1$ and $T=0$.
    
    \item Treatment applied to one user does not affect the outcome of other user (no interference).
\end{itemize}
\end{assumption}

\begin{assumption}{Unconfoundedness:} \label{assump2}
\begin{equation}
    T_i \perp \left(Y_i(0), Y_i(1)\right) | X_i,
\end{equation}
where $X_i$ is a set of the pre-treatment variables for $i$-th user, for example age, gender, country, etc.
\end{assumption}

However, sometimes the analysis based on ATE only is not enough for obtaining accurate and meaningful insights. As we shared in Section 1, we have observed  many cases where a single feature change can impact different users differently. As shown in Figure 1, the estimation of ATE is not a good measure for a heterogeneous population since it is possible for the ATE to exaggerate the treatment effect of one sub-population while underestimating the treatment effect of another sub-population. In order to study heterogeneous treatment effect, we need to consider the conditional average treatment effect, which is defined as

\begin{equation} \label{cate}
    \tau(x)=\textbf{E}\left[ Y_i(1)-Y_i(0) | X_i=x\right],
\end{equation}
where $X_i$ is a set of pre-treatment variables for $i$-th user. Obtaining accurate estimates of the conditional average treatment effect $\tau(x)$ for all values of $x$ is very useful for heterogeneous treatment effect detection, because $\tau(x)$ gives the conditional average treatment effect for the subpopulation defined by the covariates $x$. For example, if the covariate is `country', then we can partition the covariate space into countries and $\tau(x)$ is the conditional average treatment effect for the users in country $x$. If $\tau(x)$ is statistically different from the average treatment effect $\Bar{\tau}$, then the country $x$ is heterogeneous.

There is an increasing demand in rigorous analysis based on heterogeneous treatment effects (HTE), and this motivates us to develop a rigorous statistical approach for HTE detection. 

\begin{figure}[h!] \label{hte}
\centering
	\includegraphics[width=0.23\textwidth]{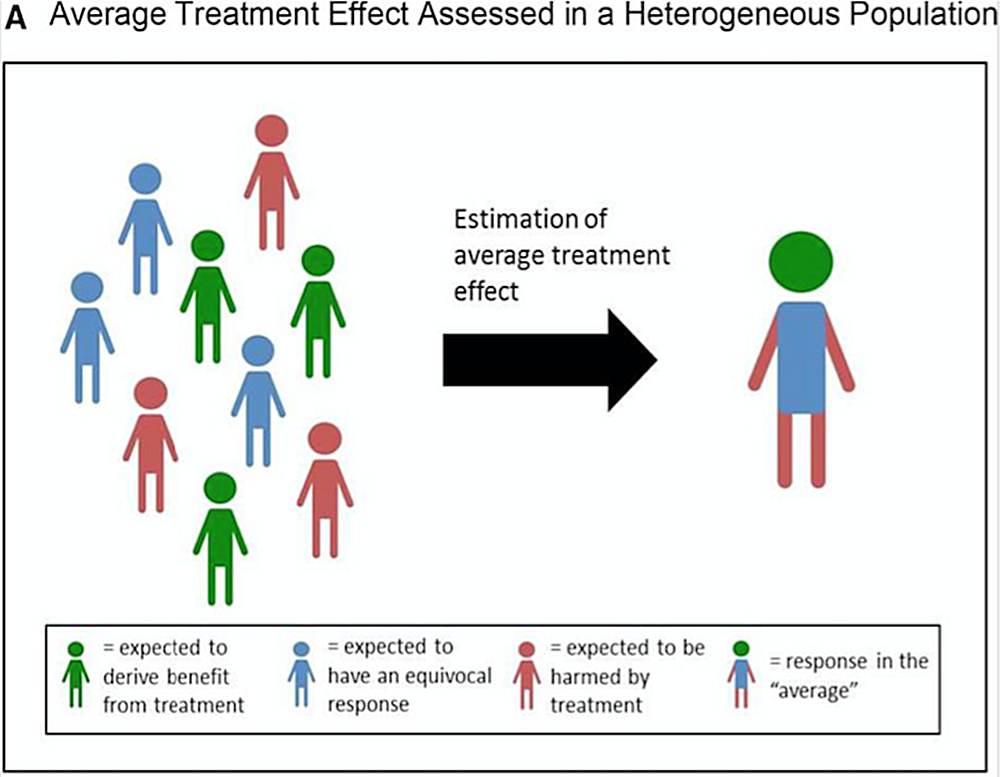} \includegraphics[width=0.23\textwidth]{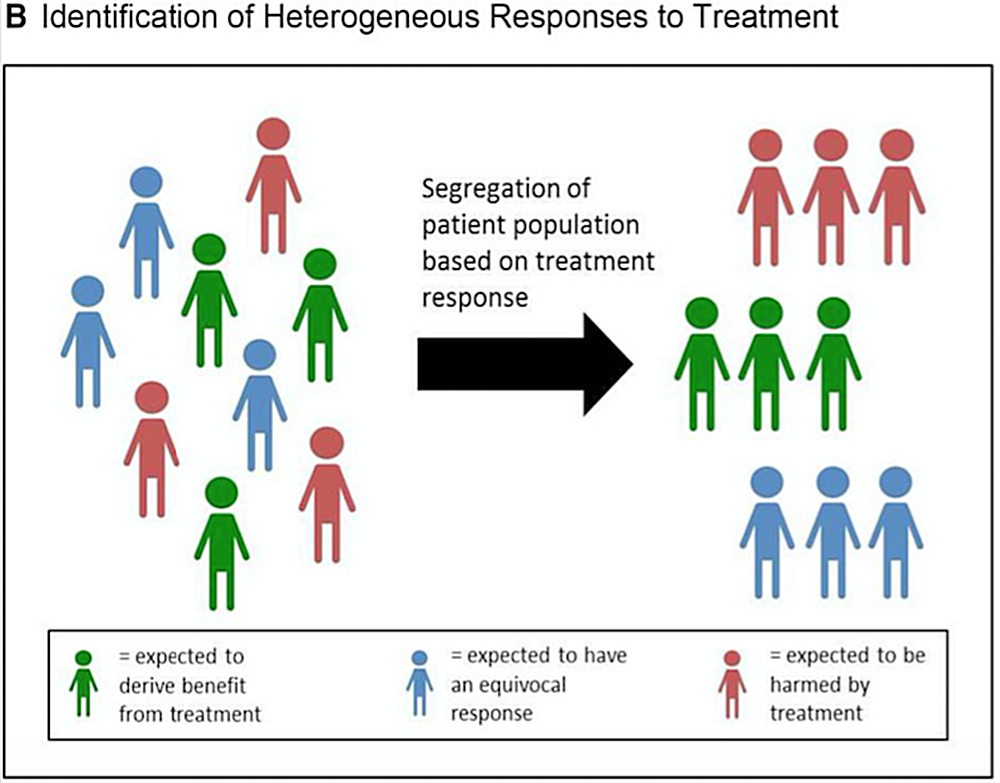}
	\caption{An illustration showing that analysis based on ATE only cannot provide accurate insights when the population is heterogeneous \citep{yeh17}.}
\end{figure}

\subsection{Naive Approaches and their Caveats}
In this section, we present some commonly used practices by practitioners which could lead to spurious discovery of HTE. Suppose that we have users from different countries and we want to find which countries have treatment effects different from ATE with respect to a metric of interest. There is a naive approach for detecting the heterogeneous countries in this problem: first run a two-sample t-test on the observations of each country and get a two-side $p$-value for each country; then choose the countries with $p$-value $<0.05$ as the result. We will refer to this approach as the ``naive approach'' from now on. 

This naive approach is straightforward and seemingly intuitive to non-statisticians. In reality, it suffers from the so-called multiple testing problem. We illustrate the problem with a simple simulation:

\begin{itemize}
    \item Step 1: Measure treatment effects for all users in 30 randomly generated subgroups from a standard Gaussian distribution so that the true ATE is zero.
    \item Step 2: Apply the naive approach and choose subgroups with $p$-values less than $0.05$ as heterogeneous.
\end{itemize}   
As Figure 2 shows, 3 out of 30 subgroups are selected as subgroups with heterogeneous treatment effect, while the truth is that the ATE estimator is 0 in this simulation and none of the subgroups is heterogeneous.  

\begin{figure}[h!] \label{multipletesting}
\centering
	\includegraphics[width=0.5\textwidth]{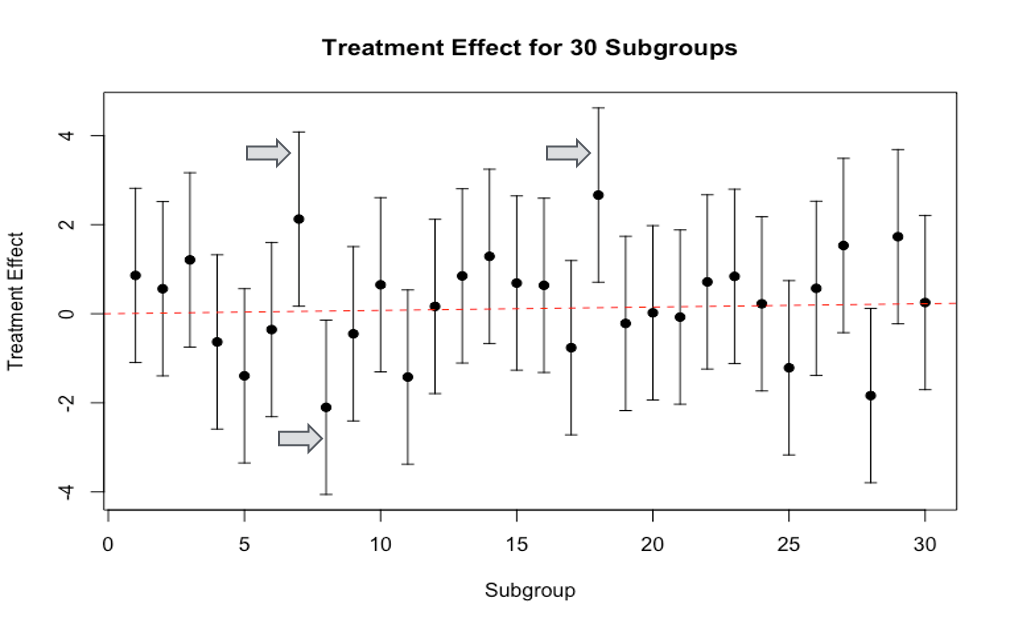} 
	\caption{A simulation example showing that the naive approach is suffering multiple testing problem. }
\end{figure}

 Bonferroni correction method \citep{bonferroni36, dunn58, dunn61} can be used to mitigate the multiple testing problem by controlling the family-wise error rate (FWER) \citep{bonferroni36}. The FWER is the probability of rejecting at least one true hypothesis. However, the Bonferroni method is known to be very conservative \citep{bh95, perneger98}, leading to a high rate of false negative and low statistical power, defined as $\mathbb{P}(\text{reject } H_0| H_1)$, where $H_0$ is the null hypothesis and $H_1$ is the alternative hypothesis. 

\section{False Discovery Rate Controlled HTE Detection}
Because of the caveats of the methods we have discussed in the previous section, in this section, we introduce the methods we propose for HTE detection, which can deal with the multiple testing problem and maintain large enough power at the same time.

In order to deal with the multiple testing problem and in the meantime reduce the conservativeness, Benjamini and Hochberg \citep{bh95} introduced the false discovery rate (FDR), which is defined as:
\begin{definition}{False Discovery Rate:}

Let $Q$ be the proportion of false positives among all the detected (rejections of the null hypothesis). Then
\begin{equation}
    \textbf{FDR}=\textbf{E}\left[Q\right].
\end{equation}
\end{definition}

In order to control FDR, we need to control the expected proportion of discoveries that are false. In addition, methods that control FDR tend to be much less conservative than the Bonferroni method. Thus in our proposed HTE detection approach, we are able to control FDR and ensure large enough power at the same time.

\subsection{Detection for Heterogeneous Subgroups}
When we run an A/B testing experiment, we are often interested in knowing which subgroups of users have treatment effects different from ATE. For example, at Snap, we have users from more than 200 countries and we are interested in finding out which countries have higher or lower treatment effects comparing with the average with respect to the metric of interest.

In this process, we need to make sure that there are not too many false discoveries in our result list. To achieve this, we choose to use the Benjamini-Hochberg (BH) procedure \citep{bh95} to control the FDR. The BH procedure is known to control FDR if the test statistics are independent or obey the positive regression dependence on a subset property introduced in \cite{by01}. It is one of the most popular FDR control methods due to its simplicity. For example, suppose that we have $p$-values from $m$ independent hypothesis testings $H_1, \ldots, H_m$ that ranks in an ascending order: $p_{(1)}, \ldots, p_{(m)}$, and we want to control FDR level at $q$. The BH procedure finds the largest $k$ such that $p_{(k)}\leq \frac{k}{m}q$, and rejects the null hypothesis for all $H_{(i)}$ for $i\leq k$. By doing so, it theoretically guarantees that the FDR is controlled under $q$.

To detect heterogeneous subgroups, we need to estimate the conditional average treatment effects defined in \eqref{cate} for the subgroups. Although the values of individual treatment effects are not available due to the fundamental problem of causal inference, we are able to construct a transformed outcome (TO) for each user as an alternative measure of individual treatment effect. Let $Y_i^{obs}$ be the observed outcome for $i$-th unit. In addition, let $p$ be the assignment probability, which, in practice, is the traffic percentage assigned to treatment group in an A/B test. The transformed outcome for the $i$-th unit, $Y_i^*$, is then defined as:
\begin{definition}{(Transformed Outcome):}
\begin{equation} \label{to}
    Y_i^*=Y_i^{obs}\times \frac{(T_i-p)}{p(1-p)}.
\end{equation}
\end{definition}

A desirable property of the TO is that under the unconfoundedness assumption the conditional expectation $\textbf{E}\left[ Y_i^* | X_i=x\right]$ equals the conditional average treatment effect $\tau(x)$ \citep{athey15}. 

We propose the following method, which combines BH method and Transformed Outcome, to detect heterogeneous subgroups. Suppose that we have $n$ users from $p$ subgroups, and we want to detect the subgroups with heterogeneous treatment effects that are different from the average treatment effect with controlled FDR. We propose the following procedure.
\textbf{HTE-BH method:}
\begin{itemize}
    \item \textbf{Step 1:} Construct a $n \times p$ design matrix $\mathbf{X}$ such that $\mathbf{X}_{i,j}=1$ if the $i$-th user belongs to the $j$-th subgroup.
    
    \item \textbf{Step 2:} Calculate the transformed outcomes $Y^*$ for all users based on the formula in Equation~\eqref{to}, and then subtract the estimated ATE $\Bar{Y}(1)-\Bar{Y}(0)$ from all transformed outcomes. Let $\mathbf{Y}$ be the vector of the resulted outcomes.
    
    \item \textbf{Step 3:} Run a linear regression using $\mathbf{Y}$ as the response and $\mathbf{X}$ as the design matrix, and get the $p$-values for the coefficient estimates corresponding to all the subgroups.
    
    \item \textbf{Step 4:} Apply the BH procedure on the $p$-values to finalize the list of selected heterogeneous subgroups.  
    
\end{itemize}

The design matrix $\mathbf{X}$ created in Step 1 is orthogonal in this case, therefore the $p$-values derived from the linear regression are independent, thus the BH procedure can control the FDR at a pre-specified level $q$. In Step 2, we subtract the estimated ATE from the transformed outcomes in order to detect the subgroups with treatment effects different from the ATE. For simplicity, we assume the estimated ATE as a parameter. Even though this overlooks the fact that the estimated ATE is a random variable, it has practical meanings as practitioners are usually interested observing which subgroups are statistically different from the observed average treatment effect over all users in an experiment.
Note that getting $p$-values from the way described in Step 3 is equivalent to getting $p$-values from running independent t-tests for all subgroups.

\subsection{Detection for Heterogeneous Factors}
Besides detecting heterogeneous subgroups, figuring out which factors contribute to the treatment effect heterogeneity is another important task of interest in practice. At Snap, we have constructed hundreds of user properties anonymously, including user demographic information such as age, gender, as well as user engagement levels, such as how users use snaps, stories or discover. Often times when we are presented with subtle results of an experiment, we don't even know which of these factors we should deep dive into. By identifying the factors contributing to the heterogeneity in treatment effect, we can more effectively dive into the corresponding factors and derive insights. The HTE-BH method is easy and simple to implement for detecting heterogeneous subgroups but is not applicable for detecting heterogeneous factors because in this case we are not able to construct an orthogonal design matrix in the Step 1 of HTE-BH method. For that reason we consider to use the `Knockoff' method \citep{barber15} in our proposal to control FDR for heterogeneous factors. 

The `Knockoff' is a recently proposed FDR control method by \cite{barber15}. Suppose that the response of interest $y$ obeys the classical linear model 
\begin{equation}
    \mathbf{y}=\mathbf{X}\mathbf{\beta} + \mathbf{\epsilon},
\end{equation}
where $\mathbf{y} \in \mathbb{R}^n$ is a vector of $y$, $\mathbf{X} \in \mathbb{R}^{n \times p}$ is any fixed design matrix, $\mathbf{\beta}$ is a vector of unknown coefficients, and $\mathbf{\epsilon} \sim \mathcal{N}(0, \sigma^2 \mathbf{I})$ is Gaussian error. Note that $n$ is the number of observations and $p$ is the number of variables. For the Knockoff method, we assume that $n\ge 2p$, which is reasonable in practice because we are likely to have more observations than variables in most A/B tests.

Let $\mathbf{\Sigma}=\mathbf{X}^T\mathbf{X}$ after normalizing $\mathbf{X}$. The `Knockoff' procedure can be summarized as the following three steps:
\begin{itemize}
    \item \textbf{Step 1:} Construct a `knockoff' matrix $\mathbf{\Tilde{X}}$ of $\mathbf{X}$ such that $\mathbf{\Tilde{X}}$ obeys the following:\\
         $\mathbf{\Tilde{X}}^T \mathbf{\Tilde{X}} = \mathbf{X}^T\mathbf{X} = \mathbf{\Sigma}$, \\
         $\mathbf{X}^T \mathbf{\Tilde{X}} = \mathbf{\Sigma} - \textbf{diag}\{\mathbf{s}\}$, \\
        where $\mathbf{s}$ is some non-negative vector that we will construct.
     
     \item \textbf{Step 2:} Compute a statistic $W_j$ for each pair $(X_j, \Tilde{X}_j)$ such that a large positive value of $W_j$ is evidence against the null hypothesis that $j$-th variable is not included in the true model. 
     
     \item \textbf{Step 3:} Calculate a data-dependent threshold $T$ such that the FDR of the knockoff selection set $\hat{S}:=\{j: W_j\ge T\}$ is less than or equal to the pre-specified level $q$. 
\end{itemize}

In our proposal, we use the equi-correlated method in \cite{barber15} to obtain the non-negative vector $\mathbf{s}$ used in the Step 1 to construct the knockoff matrix $\Tilde{X}$. The equi-correalted method suggests using $s_j=\min\{2\lambda_{min}(\mathbf{\Sigma}), 1\}$ for all $j$, where $\lambda_{min}$ is the smallest eigenvalue of $\mathbf{\Sigma}$. After getting this $\mathbf{s}$, we then construct $\Tilde{X}$ using the formula in \cite{barber15}:
\begin{equation}
    \mathbf{\Tilde{X}}=\mathbf{X}(\mathbf{I}-\mathbf{\Sigma}^{-1}\textbf{diag}\{\mathbf{s}\}) + \mathbf{\Tilde{U}} \mathbf{C},
\end{equation}
where $\mathbf{\Tilde{U}}$ is an $n \times p$ orthonormal matrix satisfying $\mathbf{\Tilde{U}}^T\mathbf{X}=\mathbf{0}$, and $\mathbf{C}$ is a Cholesky decomposition satisfying $\mathbf{C}^T\mathbf{C}=2\textbf{diag}\{\mathbf{s}\}-\textbf{diag}\{\mathbf{s}\}\mathbf{\Sigma}^{-1}\textbf{diag}\{\mathbf{s}\}$.

There are many options available for computing the statistics $W_j$'s in the Step 2 as discussed in \cite{barber15}. We choose to use Lasso to compute the statistics $W_j$'s. Let $\mathbf{X}^*=[\mathbf{X} \text{  } \mathbf{\Tilde{X}}] \in \mathbb{R}^{n \times 2p}$ be the augmented design matrix. Recall the Lasso problem:
\begin{equation}
    \textbf{minimize}_{\mathbf{\beta}} ||\mathbf{y}-\mathbf{X}^*\mathbf{\beta}||_2^2+\lambda||\mathbf{\beta}||_1.
\end{equation}

Define $Z_j=\text{sup}\{\lambda: \hat{\beta}_j(\lambda) \neq 0 \}$, that is, the largest tuning parameter $\lambda$ that first allows the $j$-th variable enters the model. Note that $(Z_j, Z_{j+p})$ is a pair correspondent to the $j$-th original variable and its knockoff. Then we calculate $W_j$ as:
\begin{equation}
    W_j = (Z_j- Z_{j+p})*\text{sign}(Z_j-Z_{j+p}),
\end{equation}
for $j=1, \ldots, p$.

Let $\mathcal{W}$ be the set $\{|W_1|, \ldots, |W_p| \} \backslash \{0\}$. In the Step 3, \cite{barber15} proposes to use the threshold 
\begin{equation}
T:=\min \left\{t \in \mathcal{W}: \frac{1+\#\{j: W_j \leq -t\}}{\max\{\#\{j: W_j \ge t \}, 1\}} \leq q  \right\}.    
\end{equation}

Theorem 2 in \cite{barber15} claims that the knockoff selection set $\hat{S}:=\{j: W_j\ge T\}$ is theoretically guaranteed to have FDR less than $q$.


We propose the following procedure to detect the variables that contribute to the heterogeneity in treatment effects while controlling FDR.

\textbf{HTE-Knockoff method:}
\begin{itemize}
    \item \textbf{Step 1:} Construct a design matrix $\mathbf{X}$ based on the set of the pre-treatment variables.  
    
    \item \textbf{Step 2:} Calculate the transformed outcomes $Y^*$ for all users based on the formula in Equation~\eqref{to}, and then subtract the estimated ATE $\Bar{Y}(1)-\Bar{Y}(0)$ from all transformed outcomes. Let $\mathbf{Y}$ be the vector of the resulted outcomes.
    
    \item \textbf{Step 3:} Create a knockoff matrix $\Tilde{\mathbf{X}}$ of $\mathbf{X}$.
    
    \item \textbf{Step 3:} Run a Lasso regression using $\mathbf{Y}$ as the response and $\mathbf{X}^*=[\mathbf{X} \text{  } \mathbf{\Tilde{X}}]$ as the design matrix.
    
    \item \textbf{Step 4:} Follow the procedure of the Knockoff method to get the knockoff selection set of the heterogeneous variables.  
    
\end{itemize}

\begin{figure*} \label{sim_gaussian}
\begin{center}
    \includegraphics[width=0.3\textwidth]{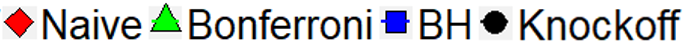}
\end{center}
    \includegraphics[width=0.45\textwidth]{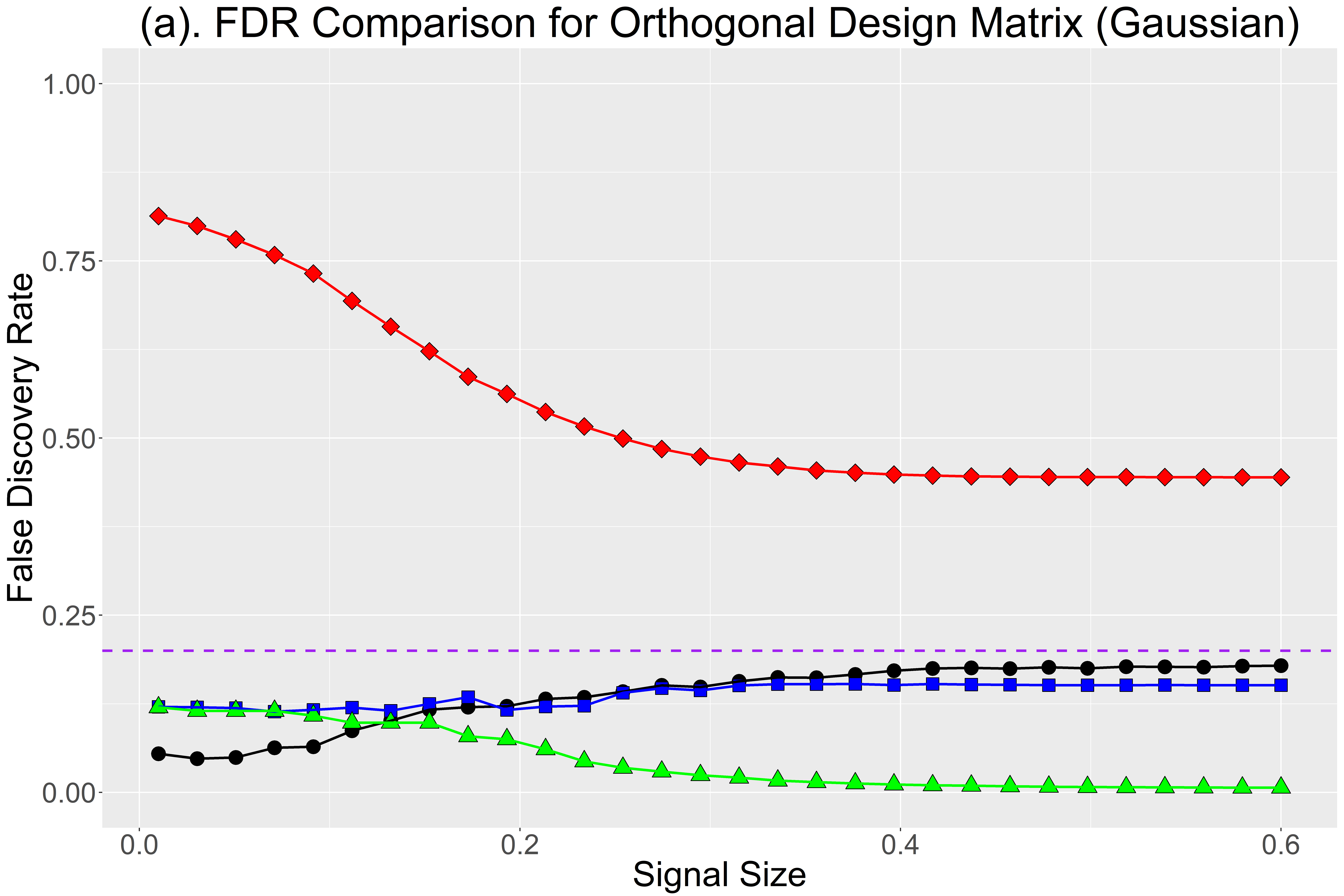} \includegraphics[width=0.45\textwidth]{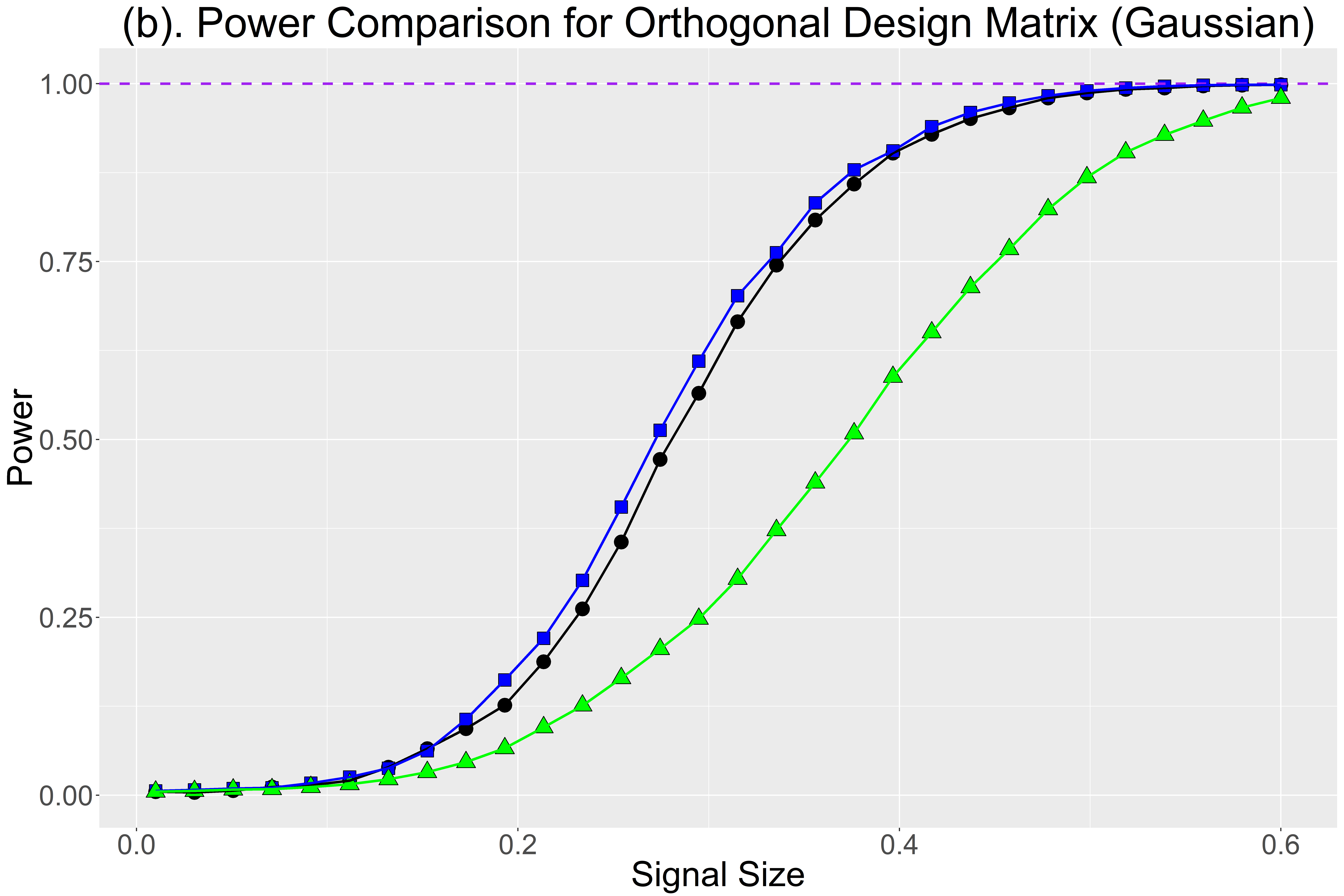}\hfill
	\includegraphics[width=0.45\textwidth]{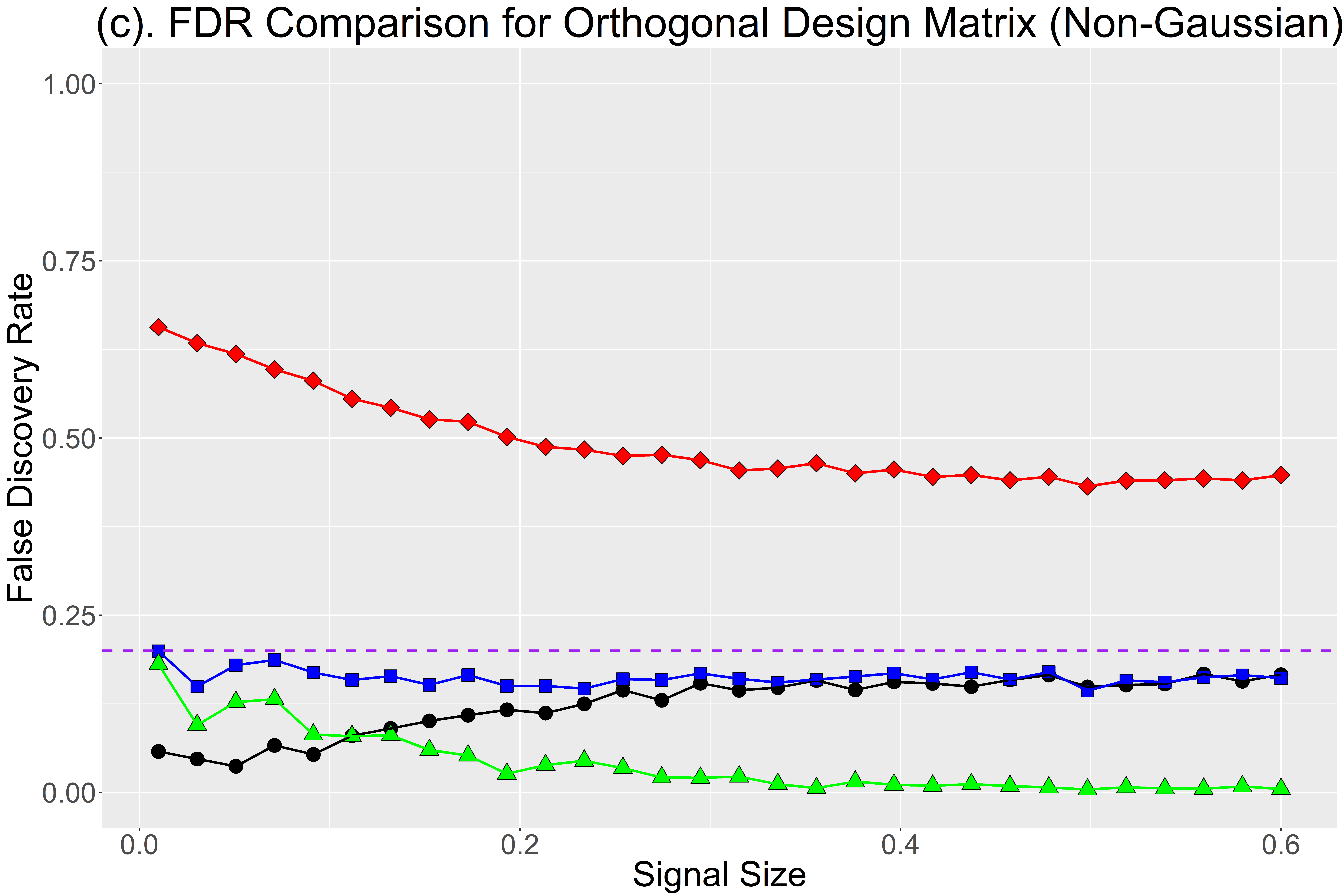} \includegraphics[width=0.45\textwidth]{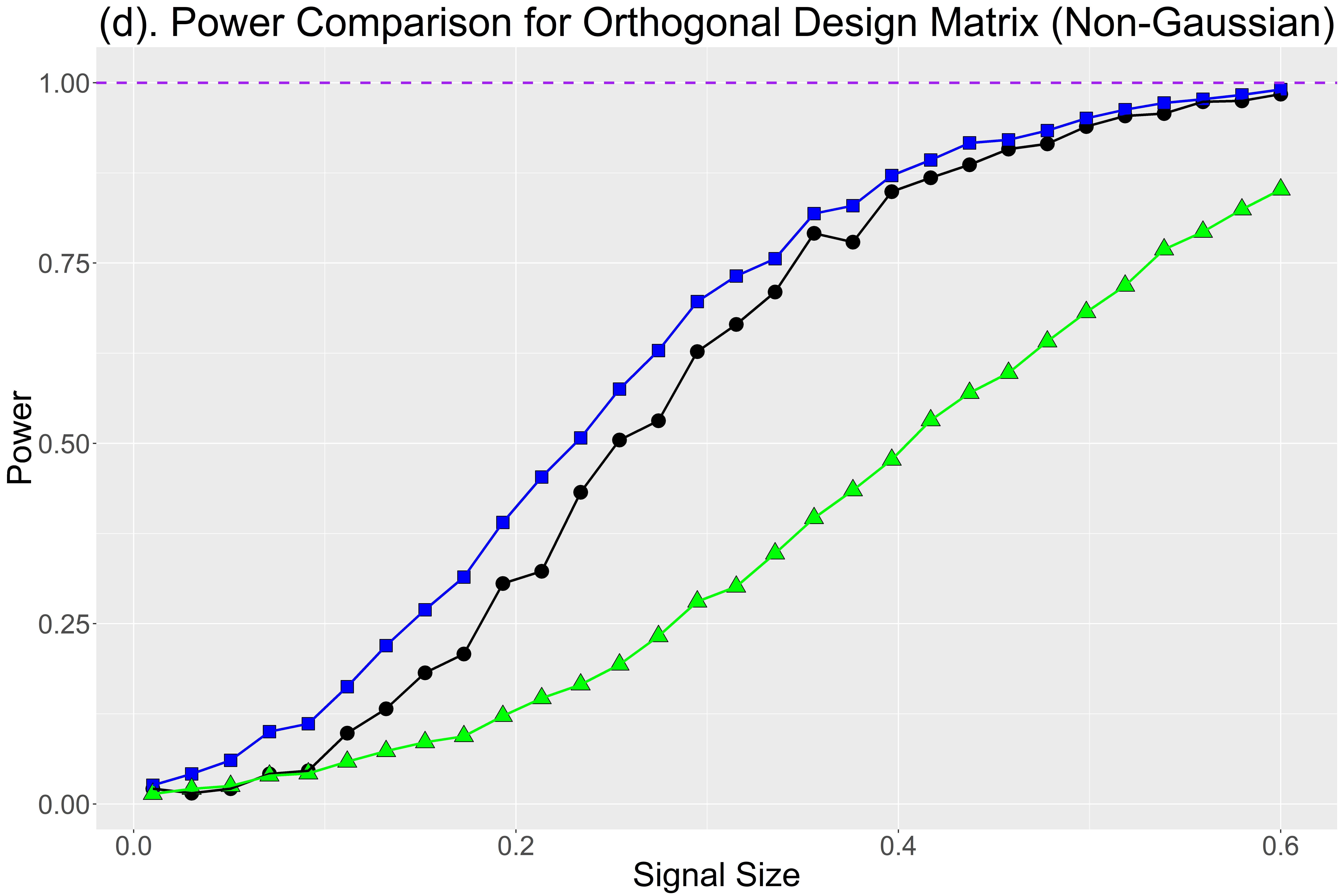}\hfill
	\includegraphics[width=0.45\textwidth]{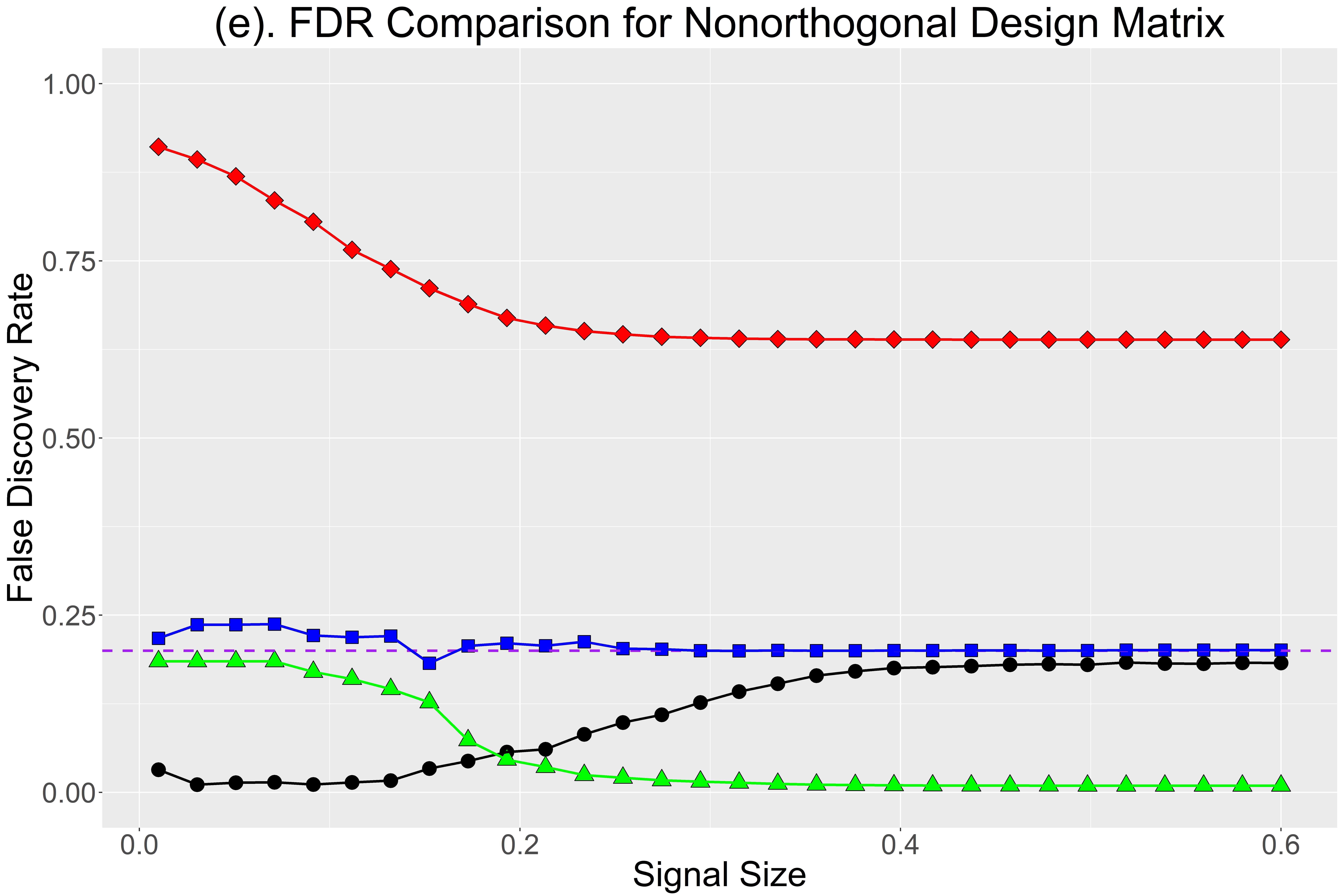} \includegraphics[width=0.45\textwidth]{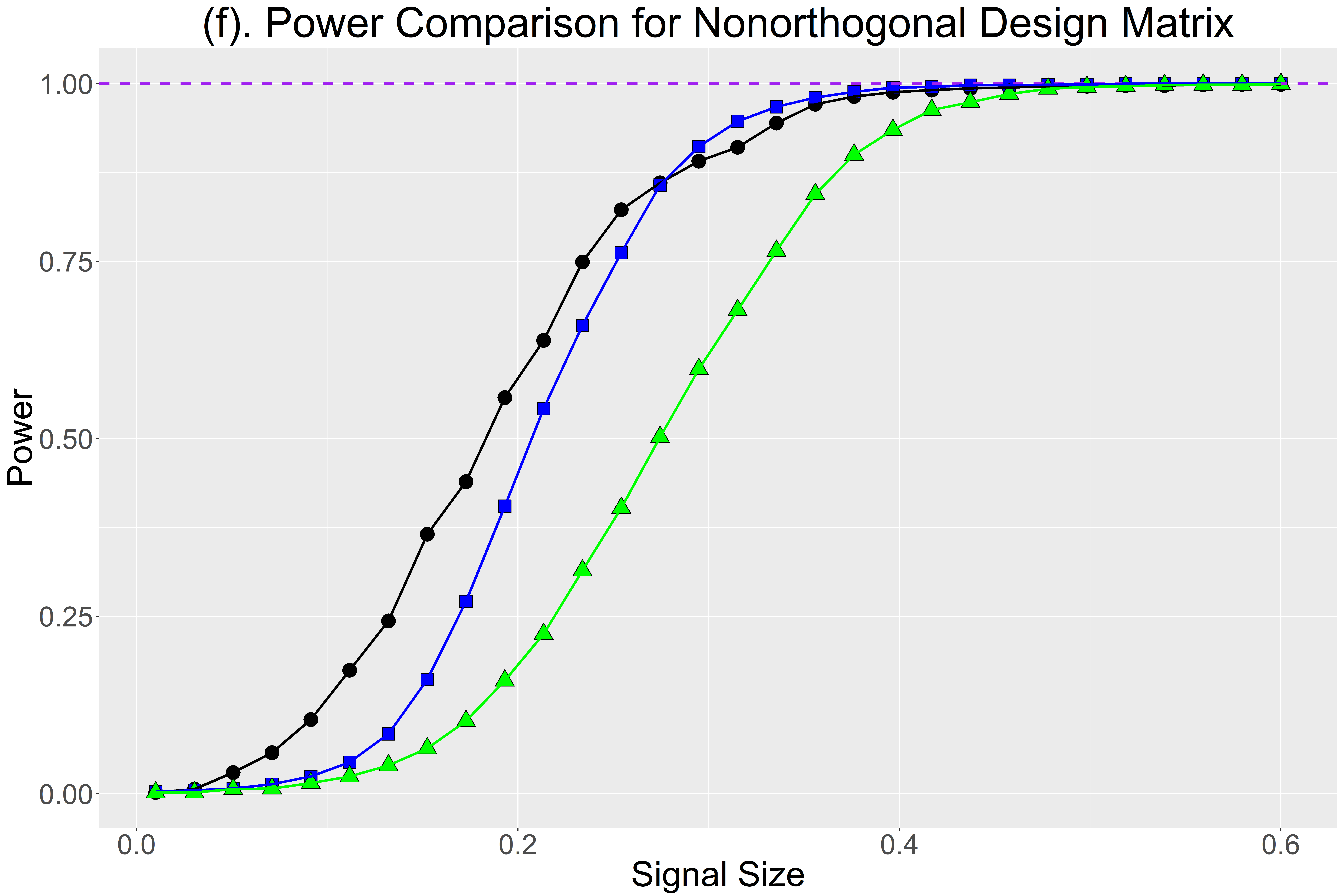}
	\caption{Simulation results for the comparison of HTE-Knockoff, HTE-BH, Bonferroni and Naive approach. For (a) and (b), the outcomes follow Gaussian distribution. For (c) and (d), the outcomes are non-Gaussian distributed. For (e) and (f), the design matrix is non-orthogonal. The results are averaged over 100 replicates. }
\end{figure*}

Note that our proposed HTE-Knockoff method can also detect heterogeneous subgroups because it works for any full-rank design matrix regardless of orthogonality. In addition, the HTE-Knockoff method is applicable when $X_i$ is a set of variables including both categorical variables and continuous variables, but we need to be careful in constructing the design matrix when there are more than one categorical variables in $X_i$. 

To lay out the problem, suppose that $X_i$ contains two categorical variables and multiple continuous variables. If we construct the design matrix to be   
\begin{equation} \label{rankdef}
\mathbf{X^*}=[\mathbf{X}_{cat1}  \text{  } \mathbf{X}_{cat2}  \text{  }  \mathbf{X}_{cont}],    
\end{equation}
where $\mathbf{X}_{cat1}$ and $\mathbf{X}_{cat2}$ are constructed to be the design matrices for subgroups defined by the two categorical variables respectively and $\mathbf{X}_{cont}$ is the design matrix created as in Step 1 of HTE-Knockoff method, then this design matrix $\mathbf{X^*}$ is rank-deficient because any column of $[\mathbf{X}_{cat1}  \text{  } \mathbf{X}_{cat2}]$ can be represented as a linear combination of the rest of the columns. 
It is not valid to apply HTE-Knockoff method using rank-deficient design matrix, since there will be infinitely many solutions for the coefficients in a regression if the design matrix is rank-deficient. So we should not create design matrix in this way. The implication is that our HTE-Knockoff method is not able to detect the heterogeneous subgroups defined by more than one categorical variable (e.g. country and mobile device model) simultaneously. The set of variables need to be composed of at most one categorical variable and simultaneously detect other continuous factors, since the design matrix $[\mathbf{X}_{cat1}  \text{  } \mathbf{X}_{cont}]$ is full-rank and can be used in HTE-Knockoff method.

Furthermore, when we have more than one categorical variable in $X_i$, a standard way to create a design matrix is to use `reference' level for the categorical variables. For example, suppose that we have two categorical variables `country' and `mobile device type'. Instead of creating a rank-deficient design matrix as in \eqref{rankdef}, we construct a design matrix with all $1$'s in the first column for the reference level where the reference level can be the subgroup for users in US and using iPhone X. The rest of the design matrix is constructed as in Step 1 of HTE-BH method except that there are no columns for the US subgroup and for the iPhone X subgroup. The resulted design matrix is full rank so that we can apply HTE-Knockoff. The interpretation of the HTE-Knockoff selecting any subgroups defined by `country' other than US is that the categorical variable `country' contributes to the heterogeneity in treatment effect. Similarly, the interpretation of selecting any subgroups defined by `mobile device type' other than iPhone X is that the categorical variable `mobile device type' is a heterogeneous factor.

The wider application of the HTE-Knockoff method is an advantage over the HTE-BH method though the later is easier to understand and implement.

\subsection{Differences between HTE-BH and HTE-Knockoff}

The HTE-Knockoff method is applicable to the case of detection for heterogeneous variables due to its validity for any fixed design matrix $\mathbf{X}$. 
 The orthogonality of design matrix in a linear regression determines the independence of the p-values. 
The non-orthogonal design matrix implies that the $p$-values from using a linear regression are not independent. Although the HTE-BH procedure can also control FDR under the positive regression dependence on a subset property \citep{by01}, this condition is generally not satisfied when the design matrix is non-orthogonal. Therefore, the HTE-BH method is not applicable for detecting variables that contributes to the heterogeneity in treatment effects.

\cite{barber15} has done a comparison between the Knockoff method and the BH procedure for the orthogonal design matrix setting. We are able to replicate their results and reach the same conclusion: as the signal varies, both methods control FDR below the pre-specified level and have almost same powers (see Figure 3 (a) and (b)), however, the empirical FDR for the HTE-Knockoff method tends to be much smaller than the HTE-BH method when the signal is so small that it is hard to detect the heterogeneous subgroups. It is actually desirable for having smaller FDR when it is hard to detect the true positives, since it saves cost by not wasting time on investigating false positives. We refer the readers to look at \cite{barber15} for more technical details about the difference in FDR between Knockoff method and BH procedure.


Figure 3 (a) and (b) present the simulation results of using our proposed methods. The signal size is the treatment effect size and it varies from very small (hard to detect) to relatively large (easy to detect). For a comparison, we also include the results for using the Naive approach and the Bonferroni method introduced in Section 2. We can see that the Naive approach fails to control the FDR, while the Bonferroni method is too conservative that it has much lower power than our proposed methods. 

The transformed outcomes generated from using real data are usually not Gaussian distributed. Therefore, we also conduct a simulation where the responses are generated from a distribution similar to the transformed outcomes we get from a real experiment. We can see from Figure 3 (c) and (d) that both HTE-BH and HTE-Knockoff still control FDR empirically, and when the underlying treatment effect size is large enough, both can detect all the positives (power=1).

When the design matrix is non-orthogonal, as Figure 3 (e) and (f) show, the HTE-BH procedure is not able to control FDR under the targeted level when signal size is relatively small while the HTE-Knockoff method still gives very small FDR.

\subsection{Empirical Results}
We apply HTE-BH and HTE-Knockoff on two real experiment datasets. 
In the first experiment, HTE-BH and HTE-Knockoff give almost same selections for heterogeneous subgroups (see Figure 4), and if we use the Naive approach, it will selects a lot more, which obviously has many false positives. The HTE result shows the drastically different effects in English speaking countries (colored points below ATE in Figure 4) versus non-English speaking countries (colored points above ATE in Figure 4), after which we retrospectively understood that the new layout of the experiment favors non-English content while suppresses high quality content in English.

\begin{figure}[h!]\label{real1}
\centering
	\includegraphics[width=0.5\textwidth]{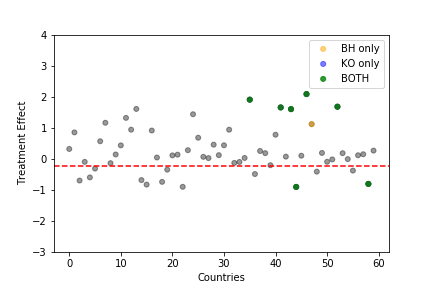}
	\caption{Real data results from using HTE-BH and HTE-Knockoff. Here we set the pre-specified FDR control level $q=0.2$. 
	\color{black}}
\end{figure}

In the second experiment, we see from Figure 5 that the HTE-BH method selects one subgroup as heterogeneous, while the HTE-Knockoff method selects none. This is very likely to be a scenario that the true treatment effects are too small to be detected, so that the HTE-Knockoff method tends to be more conservative than the HTE-BH to avoid making any false positive. This is consistent with the simulation shown in Figure 3(c).
\begin{figure}[h!] \label{real2}
\centering
	\includegraphics[width=0.5\textwidth]{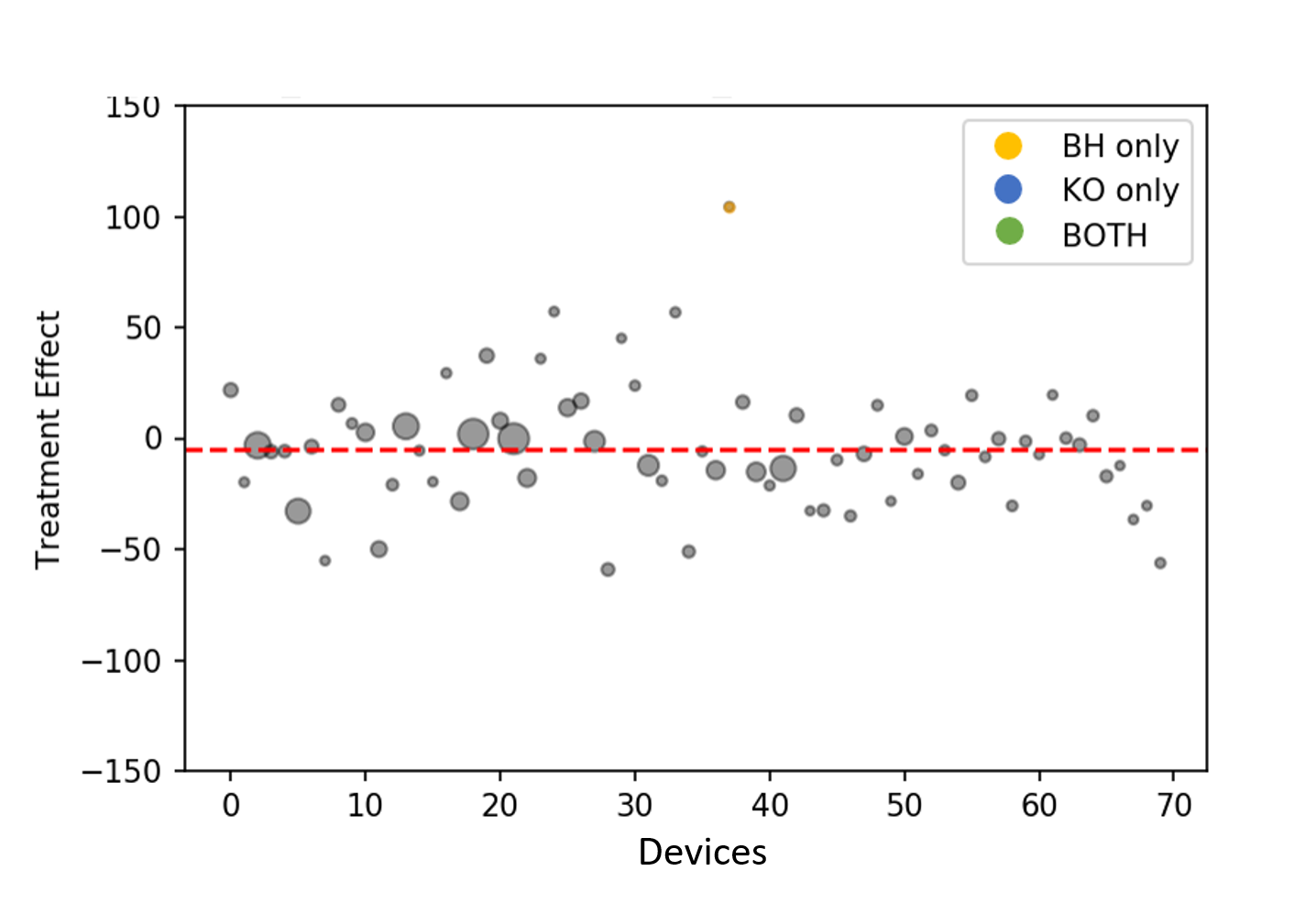}
	\caption{A real experiment example showing that the Knockoff tends to be more conservation than the BH when the signal sizes are very small, in order not to make possible false positives. }
\end{figure}

\section{Discussion}
\begin{figure*}\label{tess}
	\includegraphics[width=0.49\textwidth]{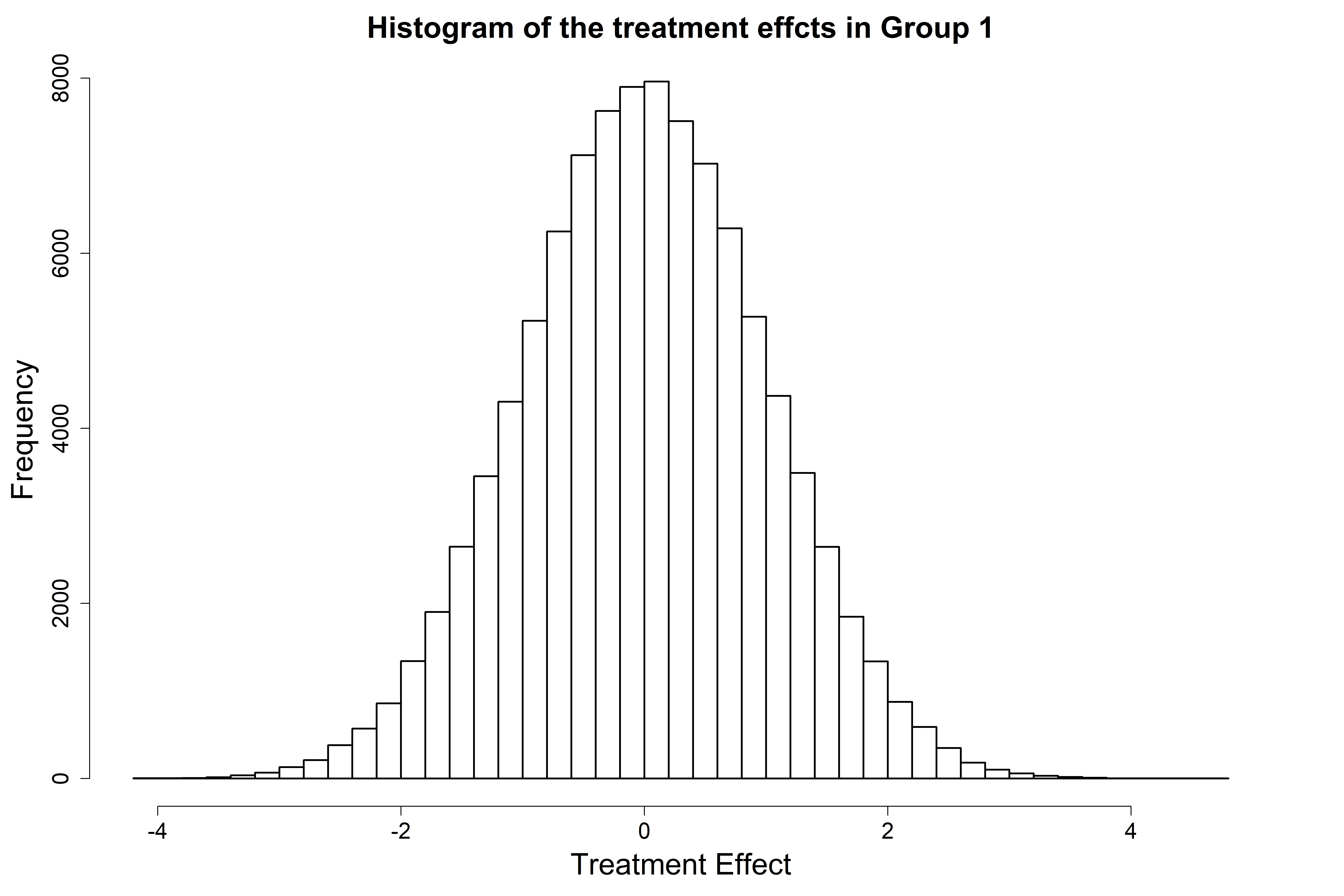} \includegraphics[width=0.49\textwidth]{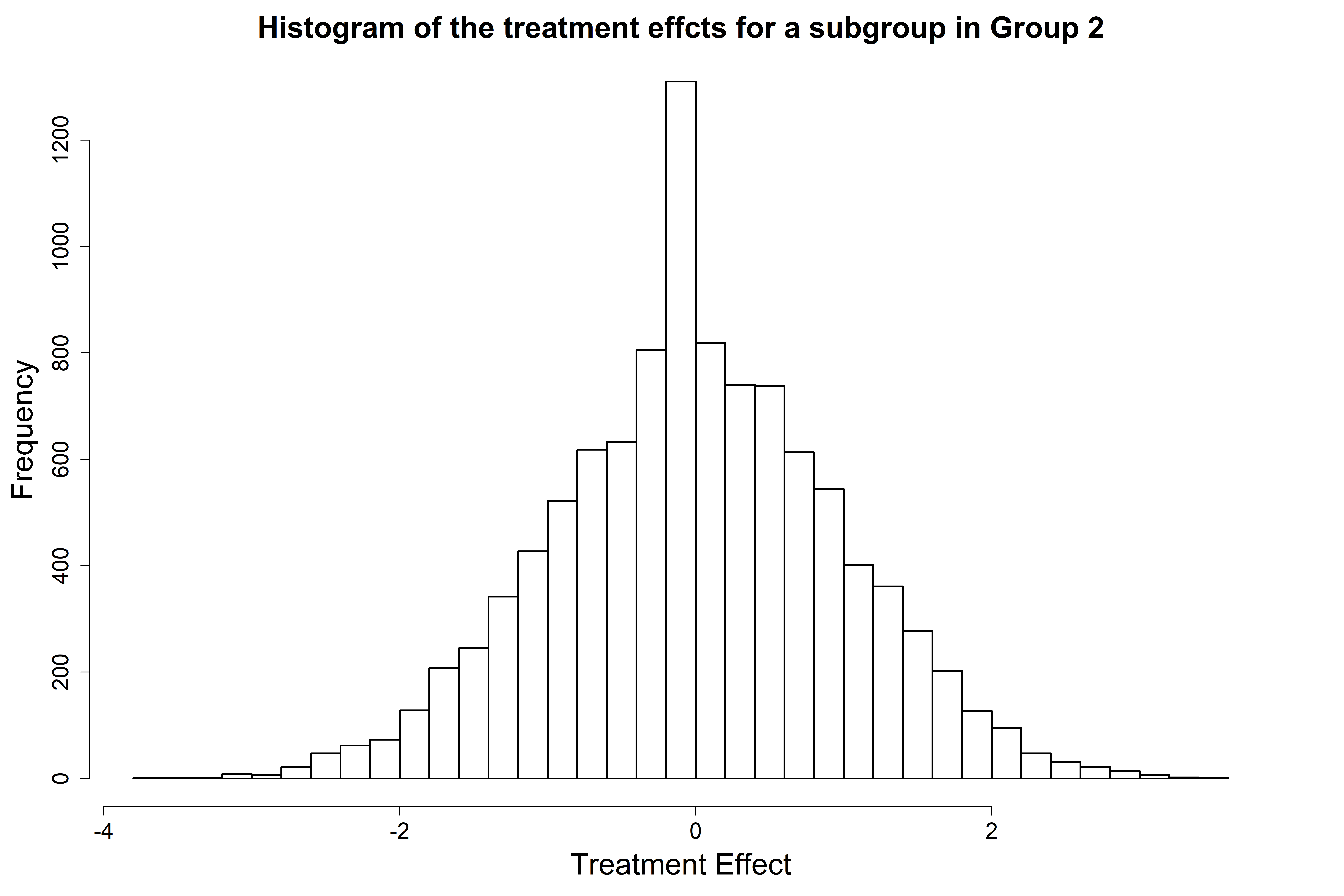}\hfill
	\caption{Left: the histogram of the treatment effects in Group 1. Right: the histogram of the treatments effects for a subgroup in Group 2. }
\end{figure*}

\subsection{Scalability}
We have thought about the scalability of our proposed methods. Instead of running the method on each experiment separately, we consider to apply either HTE-BH or HTE-Knockoff on multiple experiments using only one regression model. The first difficulty we encounter is that one unit can be in a control group in one experiment but in a treatment group in another experiment. As we notice from the formula of the transformed outcome, such unit will have two different transformed outcomes. Therefore, the transformed outcome cannot be used as a response for each unit in the regression for multiple experiments. 

Nonetheless, if the probability of treatment assignment $p$ is $0.5$, this first issue can be solved by using an alternative approach to the `transformed outcome' method, which is called the `transformed design matrix' method. As the name indicates, instead of transforming the outcomes, we transform the values in the design matrix. We then can use the observed outcome without transformation as the response and apply HTE-BH or HTE-Knockoff on the transformed design matrix.

\subsubsection{Transformed Design Matrix}
Let $p$ be the proportion of the participant assigned into the treatment group, and let $T$ be the indicator of receiving treatment or not. Then by algebra, we can show that 
\begin{equation}
\frac{1}{n} \sum_{i=1}^{n} (Y_i^*-\mathbf{X}_i\mathbf{\beta} )^2 = \frac{1}{n}\sum_{i=1}^{n} w_i \left(Y_i^{obs}-\mathbf{X}_i\mathbf{\beta}  (T_i-p)V_i \right)^2,    
\end{equation}
 where $\mathbf{\beta}$ is a vector of unknown coefficients, $\mathbf{X}_i^T$ is the $i$-th row of the design matrix, $w_i=\frac{1}{(1-p)^2}*\mathbf{1}_{\{T_i=0\}}+\frac{1}{p^2}*\mathbf{1}_{\{T_i=1\}}$, and $V_i=\frac{p}{1-p}*\mathbf{1}_{\{T_i=0\}} + \frac{1-p}{p}*\mathbf{1}_{\{T_i=1\}}$. When $p=0.5$, we can reduce it to a simple equation 
 \begin{equation}
 \frac{1}{n} \sum_{i=1}^{n} (Y_i^*-\mathbf{X}_i\mathbf{\beta} )^2 = \frac{4}{n}\sum_{i=1}^{n} \left(Y_i^{obs}-\mathbf{X}_i\mathbf{\beta}(T_i-0.5)\right)^2.     
 \end{equation}
 It implies that running a linear regression using the transformed outcome $Y^*$ as response and $\mathbf{X}$ as design matrix is equivalent to solving a linear regression problem using the observed outcome $Y^{obs}$ as response and $\mathbf{Z}$ as design matrix, where $\mathbf{Z}_i=\mathbf{X}_i(T_i-0.5)$ is the transformed design matrix. We note that \cite{tian14} has derived a similar result for the case of $p=0.5$. We refer readers to look at the details about the derivation of this special case result in their paper. Our generalized result can be derived using similar technique. 

\subsubsection{Computational Complexity Comparison}
The use of the transformed design matrix enables us to run a linear regression using $Y^{obs}$ as response and an augmented design matrix $\mathbf{Z}^*=[\mathbf{Z}_1 \ldots \mathbf{Z}_m]$ for $m$ experiments together. However, it does not improve the scalability. The computational complexity of running a linear regression is $\mathcal{O}(p^2(n+p))$ if we have $n$ observations and $p$ columns of $\mathbf{Z}_k$ for each $k=1, \ldots, m$. Therefore, the computational complexity of running our proposed HTE-BH or HTE-Knockoff method $m$ times is $\mathcal{O}(mp^2(n+p))$, while the computational complexity of running the transformed design matrix method on $m$ experiments together is $\mathcal{O}((mp)^2(n+mp))$. Unfortunately, $\mathcal{O}((mp)^2(n+mp))$ is bigger than $\mathcal{O}(mp^2(n+p))$. 

We leave the scalability problem to future research work.

\subsubsection{Advice for choices of $n$ and $p$}

We have tested the scalability of our methods using different combinations of $n$ and $p$. We implemented our methods in a toolkit at Snap and ran this toolkit on a data set with $n=24$ million and $p=20$ and a data set with $n=7$ million and $p=80$. In both cases, the task was finished within one hour on Apple 15" MacBook Pro 2.8GHz Intel Core i7. We believe that the running time can be much shorter on a more powerful system.

\subsection{Heterogeneity inherent in population v.s. Heterogeneity in treatment effect}


Besides the two methods we propose in this paper, we have also considered another way of detecting heterogeneous subgroups. The procedure is as follows:
\begin{itemize}
    \item \textbf{Step 1:} Calculate the transformed outcomes for all units and split all the units randomly into Group 1 and Group 2.
    
    \item \textbf{Step 2:} Use the transformed outcomes of the units in Group 1 to construct an empirical distribution of the transformed outcomes for the whole population.
    
    \item \textbf{Step 3:} Divide the units in Group 2 into subgroups, e.g. by countries.
    
    \item \textbf{Step 4:} Compute a degree of the anomalous pattern for each subgroup's transformed outcome distribution, comparing with the empirical distribution for the whole population in Step 2.
    
    \item \textbf{Step 5:} Rank the anomalous pattern degrees and select the most abnormal subgroups as heterogeneous.
\end{itemize}


There are many methods available to calculate sensible anomalous pattern degrees in Step 4. We choose to use the Higher Criticism \citep{donoho04, mcfowland13a}. However, we realize two issues of using this HTE detection procedure: the first issue is that we cannot control FDR using the degrees of the anomalous pattern, because we are not able to obtain a threshold value for the anomalous pattern degrees as in BH and Knockoff method; the second issue is that this method identifies the heterogeneity inherent in the population instead of the heterogeneity in the treatment effects. The second issue can be illustrated in an example in Figure 6. The left plot shows the distribution of the treatment effect values for the whole population, and the right plot shows the distribution of the values for one subgroup. If we use two-sample t-test to compare these two sets of values, the $p$-value is very large, indicating that the mean difference between the two groups is not statistically different from 0. However, if we use the Higher Criticism, we get a very large degree of anomalous pattern due to the fact that there are more zeros in the right plot than in the left plot. Therefore, this method detects a different type of heterogeneity: the heterogeneity in the distribution of outcomes for a subgroup, instead of the heterogeneity in the treatment effects for a subgroup. This type of heterogeneity is inherent in population and is not affected by treatment.

\section{Literature Review}
%
%
%
%
%
%
%
%
%
%
\subsection{Average Treatment Effects}

The Rubin causal model (RCM) was first proposed in \cite{holland86} as a statistical framework for causal inference. Based on the RCM, many statistical methods have been developed for causal inference and most of them focus on the estimation and inference on average treatment effects (ATE). There are matching methods in causal inference for estimating ATE \citep{sekhon07, pearl00}, and there also exist methods for identifying casual effects and estimating ATE using instrumental variables \citep{angrist96, aronow13}. Recently, researchers even have developed globally efficient estimators for ATE in observational studies \citep{chan16}. In addition, many digital experimentation works in industry have focused on the analysis of ATE \citep{xu16, kohavi13, johari15, deng15}.

\subsection{Heterogeneous Treatment Effects}

Recently, many researchers have shifted their attention from the ATE to the heterogeneous treatment effects (HTE). In this section we review existing related work on HTE. It is commonly understood by the researchers that a treatment having positive effect on one subgroup of people may have negative effect on another subgroup, due to the heterogeneity of the population. Therefore, it is crucial to ascertain subgroups for which a treatment is harmful or beneficial. In contrast to the ATE, the HTE is able to tell the A/B test designers about the treatment effects for subpopulations. The study for HTE has gained a lot attention over the last couple of years, which leads to many insightful ideas about learning the HTE. 

The work in \cite{imai13} estimates the treatment effect heterogeneity in a randomized evaluation program by using Squared Loss Support Vector Machine (L2-SVM) with L1 penalty (LASSO). The key part of their approach is to put two separate L1 penalties on the coefficient estimates for the pre-treatment covariates and the coefficient estimates for the interaction between treatment and pre-treatment covariates. The intuition behind it is that the interactions, which are related to causal heterogeneity, usually have weaker predictive power than the pre-treatment covariates in the model. If the coefficient estimate for the interaction term between the treatment and a pre-treatment covariate is non-zero, then this pre-treatment covariate is selected as a variable contributing to the treatment effect heterogeneity. This method distinguishes between the estimation of treatment effects and the estimation of the impact of other pre-treatment covariates of units, however, it is only applicable for randomized trials.

\cite{deng16} proposes a total variation regularized regression model to understand the structure of the HTE. In their regression model, in addition to the use of LASSO for selecting potential covariates, they further include a total variation penalty to encourage block-wise structure for the non-zero coefficients of the potential covariates. The advantage of this is that the model usually results in actionable and interpretable conclusions which are well-suited to practitioners.

\cite{athey15} uses machine learning methods to estimate heterogeneous treatment effects which are applicable to the data in both randomized trials and observational studies. They introduce the idea of transforming the observed outcomes and using the transformed outcomes in machine learning models. We put more details about the transformed outcomes in next section since it is also related to our proposal. Based on the transformed outcomes, \cite{athey15} introduces several Out-of-Sample Goodness-of-fit measures and In-Sample Goodness-of-fit measures, and then propose the Transformed Outcome Tree (TOT) method which uses regression tree model based on the Goodness-of-fit measures. They show in simulations and empirical datasets that the TOT achieves better estimation on heterogeneous treatment effects than many standard methods. However, this work does not focus on identifying subgroups that differ from average users. 

\cite{wager17} develops a causal forest method based on the idea of the causal tree model in \cite{athey15} and the extension of the well-known random forest algorithms proposed in \cite{breiman01}. This paper is, to our best knowledge, the first one to conduct both the estimation and the inference on heterogeneous treatment effects. However, this work also focuses on estimation inference on the HTE instead of understanding systematically which subgroup do/do not differ from average users. 

Besides, \cite{berry16} develops a two-stage method for finding HTE, and \cite{peysakhovich16} proposes to combine observational and experimental data to study HTE.

Although the above-mentioned works have contributed a lot into the study of estimating the HTE or drawing inference from the HTE, to the best of our knowledge, there is no literature trying to deal with the potential multiple testing problem when we conduct analysis for the HTE. For example, using L1 penalty in a regression model may help us to select a set of variables that potentially causes the heterogeneity in treatment effect, but it is possible that a large proportion of the selected variables are false positives. It is important to solve the multiple testing problem, otherwise the reproducibility of result will be low. 

\subsection{False Discovery Rate}
In our proposal, we focus on controlling the false discovery rate (FDR) in order to deal with the multiple testing problem. The concept of FDR was first introduced by Benjamini and Hochberg in \cite{bh95} and they propose Benjamini and Hochberg (BH) procedure to control the FDR level under the asusmption that the test statistics are independent. \cite{by01} extends the BH procedure to BH-Y procedure which allows the test statistics to have positive dependency. \cite{bky06} later develop a two-stage method for FDR control under positive dependency, and they argue that this is the best option when the degree of dependence is unknown.

Recently, a novel method called `Knockoff' is proposed in \cite{barber15} to control FDR. The Knockoff approach serves as a variable selection and simultaneously control the FDR during the selection procedure.

\section{Summary and Future Work}
In this paper, we propose the HTE-BH method for detecting heterogeneous subgroups with treatment effects different from the average, and propose the HTE-Knockoff method for detecting factors contributing to the heterogeneity in treatment effects. While the HTE-BH method is easier to implement, the HTE-Knockoff has wider application as it can also be used to detect heterogeneous factors. Our proposed methods have good power for detection and in the meantime deal with the multiple testing problem by controlling FDR level. 

In spite of their wide application scenarios, our current methods still have some limitations and thus can be improved in future research work.

The first limitation of our approach is the assumption of the true model being a linear regression model with Gaussian error; the theoretical properties of the original Knockoff method in \cite{barber15}
are based on this assumption. Although we show in Figure 3 (c) and (d) that the Knockoff method can still work well in FDR control in some non-Gaussian error cases, it lacks theoretical proof for such robustness. In addition, sometimes the true relationship between the treatment effect and the variables are not linear, therefore, using a linear regression may not be appropriate. Very recently, \cite{candes17} has proposed a model-free knockoff method, which, under some conditions, can work on any kind of non-linear models. This idea can be useful if we want to extend the HTE-Knockoff procedure to a more generalized setting in future work.

Another unsolved problem is the scalability. We have tried the idea of transformed design matrix to conduct HTE detection on multiple experiments, but it turns out that the computational complexity even increases. This problem is worth for further study because most companies have a large number of A/B test results available and it is not feasible for them to apply the HTE detection method on the experiments one by one.



\section{Acknowledgement}
We thank the anonymous reviewers and the editor for helpful comments on this work. We thank Dr. Gary Chan (UW Seattle) for insightful discussions on transformed outcome and Knockoff procedure. We thank the engineering team at Snap for providing valuable data sets of online controlled experiments.

\bibliographystyle{ACM-Reference-Format}
\bibliography{ref}

\end{document}